\documentclass[11pt]{article}

\usepackage[final]{acl}
\usepackage{stfloats}
\usepackage{float}
\usepackage{times}
\usepackage{latexsym}

\usepackage[T1]{fontenc}

\usepackage[utf8]{inputenc}

\usepackage{microtype}

\usepackage{inconsolata}

\usepackage{graphicx}

\usepackage{epsfig}
\usepackage{multirow}
\usepackage{amsmath}
\usepackage{amsfonts}

\def\BibTeX{{\rm B\kern-.05em{\sc i\kern-.025em b}\kern-.08em
    T\kern-.1667em\lower.7ex\hbox{E}\kern-.125emX}}

\begin{document}
\title{Unifying Streaming and Non-streaming Zipformer-based ASR}

\author{
  Bidisha Sharma$^{1}$ \quad Karthik Pandia Durai$^{1}$ \quad Shankar Venkatesan$^{1}$ \quad Jeena J Prakash$^{1}$ \\
  \bf{Shashi Kumar$^{2}$} \quad \bf{Malolan Chetlur$^{1}$} \quad \bf{Andreas Stolcke$^{1}$} \\
  $^{1}$Uniphore,  India \& USA, \quad $^{2}$Idiap Research Institute, Switzerland \\
  \texttt{bidisha@uniphore.com, karthikpandia@uniphore.com, shankar.venkatesan@uniphore.com} \\
  \texttt{jeena@uniphore.com, malolan.chetlur@uniphore.com, andreas.stolcke@uniphore.com} \\
  \texttt{shashi.kumar@idiap.ch}
}

\maketitle
\begin{abstract}
There has been increasing interest in unifying streaming and non-streaming automatic speech recognition (ASR) models to reduce development, training, and deployment costs. We present a unified framework that trains a single end-to-end ASR model for both streaming and non-streaming applications, leveraging future context information. We propose to use dynamic right-context through the chunked attention masking in the training of zipformer-based ASR models. We demonstrate that using right-context is more effective in zipformer models compared to other conformer models due to its multi-scale nature. We analyze the effect of varying the number of right-context frames on accuracy and latency of the streaming ASR models. We use Librispeech and large in-house conversational datasets to train different versions of streaming and non-streaming models and evaluate them in a production grade server-client setup across diverse testsets of different domains. The proposed strategy reduces word error by relative 7.9\% with a small degradation in user-perceived latency. By adding more right-context frames, we are able to achieve streaming performance close to that of non-streaming models. Our approach also allows flexible control of the latency-accuracy tradeoff according to customers requirements.


\end{abstract}

\section{Introduction}
In recent times, end-to-end (E2E) ASR models have started taking the main stage in industrial use-cases~\cite{povey16_interspeech}. Recurrent neural networks (RNNs) are crucial as they can model the temporal dependencies in audio sequences effectively~\cite{chiu2018state,rao2017exploring,sainath2020streaming}. The transformer architecture with self-attention has gained substantial attention in ASR to capture long distance global context and show high training efficiency~\cite{zhang2020transformer, vaswani2017attention, hsu2021hubert, chen2022wavlm}. Alternatively, ASR based on convolutional neural networks (CNNs) has also been successful due to its ability to exploit local information \cite{li2019jasper, han2020contextnet, abdel2014convolutional}. Recently, the conformer ASR model~\cite{gulati20_interspeech} was proposed for combining the advantages of CNN and transformer models, to extract both local and global information from a speech sequence~\cite{han20_interspeech, shi2021emformer, kim2022squeezeformer, yao2023zipformer}. Zipformer~\cite{yao2023zipformer} is an extension of the previous conformer models, providing a transformer that is faster, more memory-efficient, and better-performing.

Latency-accuracy is a critical trade-off for an ASR model, especially for streaming ASR models. In systems with concurrent call processing, it becomes critical to find the optimal operating point in the latency-concurrency-accuracy trio. Streaming decoders work on chunk-based processing, where, for each frame the encoder has access to, the entire left-context and a variable right-context depending on the frame's position in a chunk are used. 

Right context has a significant role in the context of a unified model in streaming and in offline production environment. Typically, the WER of the offline model is significantly lower compared to that of a streaming model. Therefore, separate models are generally trained for offline and streaming use-cases. This requires twice the compute resource to train the models and additional resource to maintain and update the models. Adding right-context helps bridge the gap in WER between offline and streaming models with a small degradation in latency in the streaming case.


In~\citet{swietojanski2023variable}, authors use variable attention masking in a transformer transducer setting, however the influence of different numbers of right-context frames is not explored and the work instead focuses on using right-context ranging from multiple chunks to full context, which may not be possible for a streaming setup. \citet{li2023dynamic} propose a dynamic chunk-based convolution, where the core idea is to restrict the convolution at chunk boundaries so that it does not have access to any future context and resembles the inference scenario. Our approach, by contrast, uses limited additional right-context frames beyond chunk boundaries. Our proposed method is also different from that of \citet{tripathi2020transformer}, where initial layers are trained with zero right context and the final few layers are trained with variable context. If we wanted a streaming model with different latency during inference, the model would need to be retrained. \citet{zhang2020unified} use dynamic chunk sizes for different batches in training and
the attention scope varies from left-context only to full context. The authors in~\citet{wu2021u2} further enhance their strategy by employing bidirectional decoders in both forward and backward direction of the labeling sequence. In both passes, they use either full right-context or full left-context attention masking, which may adversely impact the real-time streaming use-case.



Our work is significantly different from the aforementioned approaches in terms of training with variable right-context while decoding with extra right-context frames in addition to the chunk being decoded in the inference phase. 
We propose to unify streaming and non-streaming zipformer-based ASR models by leveraging future context. The conventional zipformer model uses chunked attention masking and utilizes only left-context while we use a variable number of right-context frames for different mini-batches during training, providing the flexibility to select a desired number of right-context frames during inference, according to the desired accuracy-latency tradeoff. We study the effect of choosing different amounts of right context on latency and accuracy, finding that as the number of decoding right-context frames increases, the streaming zipformer ASR model can approach the performance of the corresponding non-streaming model without significantly degrading latency. We evaluate our method on both open-source read speech and industry-scale production-specific conversational speech data.

\vspace{-0.2cm}
\section{Right-context in Zipformer}
\label{sec:Mask_zipfromer}
Here we review the zipformer model and the attention masking employed to incorporate right-context information~\cite{gulati20_interspeech,yao2023zipformer}.

\subsection{Zipformer model}\label{sec:Zipformer}
\begin{figure}[t]
 \centering
\centerline{\epsfig{figure=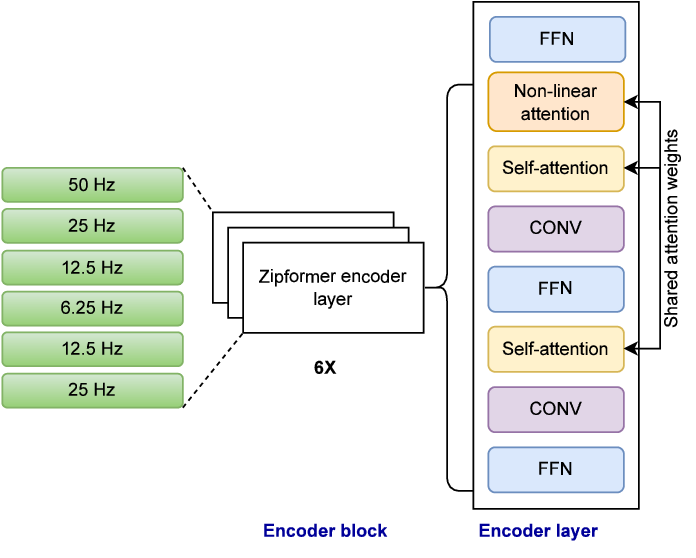,scale=0.5}}
{\caption{Zipformer encoder architecture showing each layer at different frame rates (left) and different modules in each encoder layer (right). }
\label{fig:zipformer}}
\end{figure}

The zipformer model is a significant advancement in transformer-based ASR encoding, offering superior speed, memory efficiency, and performance compared to conventional conformer models. A conformer model adds a convolution module to a transformer to add both local and global dependencies. In contrast to the fixed frame rate of 25Hz used by conformers, the zipformer employs a U-Net-like structure, enabling it to learn temporal representations at multiple resolutions in a more streamlined manner. 

In the zipformer encoder architecture, we have six encoder blocks, each at different sampling rates learning temporal representation at different resolutions in a more efficient way. Specifically, given the acoustic features with frame rate of 100 Hz, a convolution based module reduces it first to 50 Hz, followed by the six cascaded stacks to learn temporal representation at frame rates of 50Hz, 25Hz, 12.5Hz, 6.25Hz, 12.5Hz, and 25Hz, respectively as shown on the left side of Figure~\ref{fig:zipformer}. The middle block operates at 6.25 Hz undergoing stronger downsampling, thus facilitating more efficient training by reducing the number of frames to process. The frame rate between each block is consistently 50 Hz. Different stacks have different
embedding dimensions, and the middle stacks have larger dimensions. The output of each stack is
truncated or padded with zeros to match the dimension of the next stack. The final encoder output
dimension is set to the maximum of all stacks' dimensions.

The inner structure of each encoder block is shown in the right side of Figure~\ref{fig:zipformer}. The primary motivation is to reuse attention weights to improve efficiency in both time and memory. The block input is first processed by a multi-head attention module, which computes the attention weights. These weights are then shared across a non-linear attention module and two self-attention modules. Meanwhile, the block input is also fed into a feed-forward module followed by the non-linear attention module.

\subsection{Attention masking}
The multi-head self-attention facilitates fine-grained control over neighboring information at each time step. At each time t, $Zipformer(x, t)$ may be derived from an arbitrary subset of features in x, as defined by the masking strategy implemented in the self-attention layers~\cite{vaswani2017attention}. Given the attention input $Y = (y_1, \ldots , y_{L_y} ), y_t \in \mathbb{R}$  self-attention computes 
\begin{equation}\label{Intent_loss}
Q=\mathcal{F}^q(Y), K=\mathcal{F}^v(Y), V=\mathcal{F}^v(Y),\\
\end{equation}
\begin{equation}\label{mask}
Att(Q,K,V) = softmax\left(\frac{\mathcal{M}(QK^T)}{\sqrt d}\right)V^T,\\
\end{equation}
where, $d$ is the attention dimension, $\mathcal{M}$ is the attention mask with values 0 and 1 of dimension $L_y \times L_k$. The attention mask in the Equation~\ref{mask} regulates the allowance of number of left and right-context frames corresponding to each frame of $Y$.

\subsection{Right-context attention masking}\label{sec:attention_mask}
The attention masks constrain the receptive field in each layer without the need for physically segmenting the input sequence. In a streaming ASR setup, to mitigate computational costs and latency, the processing occurs at the chunk level rather than at the frame level. A specific number of frames are grouped into chunks, and each chunk is then encoded as a batch. Following \citet{shi2021emformer,chen2021developing}, we use chunked attention masking to confine the receptive field during self-attention computation. In conventional chunked attention masking, each frame within a chunk is exposed to varying extents of left- and right-context frames. The initial frames in a chunk have access to some right-context frames, while the later frames have no access to right-context frames, enforcing a causal constraint at chunk boundaries.

\begin{figure}[ht]
\centering
\centerline{\epsfig{figure=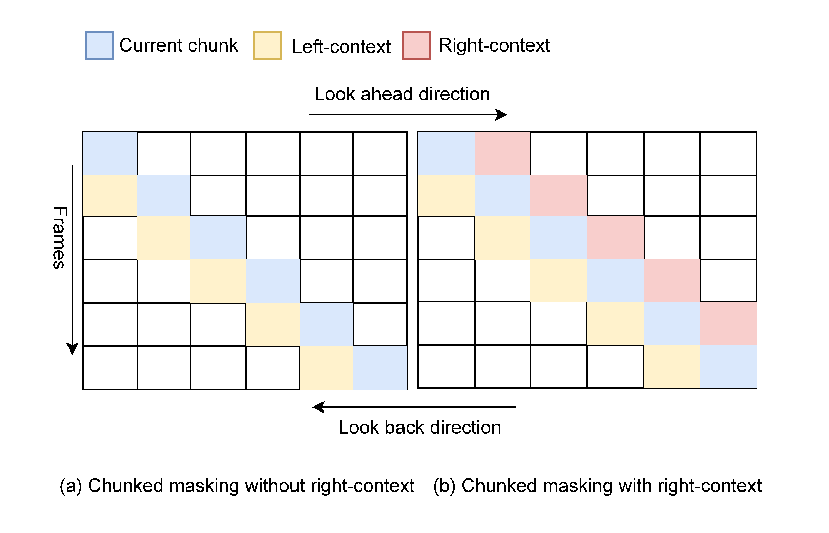,scale=0.5}}
{\caption{Attention masking in zipformer; (a) chunked masking with left-context and no right-context, (b) chunked masking with both left-context and right-context.}
\label{fig:mask}}
\end{figure}

The conformer and zipformer ASR recipes in {\it k2-fsa icefall}\footnote{https://github.com/k2-fsa/icefall}~\cite{gulati20_interspeech,shi2021emformer} deploy chunked attention masking and use only left-context as shown in Figure~\ref{fig:mask}(a). For streaming decoders, each frame in the encoder accesses left-context and variable right-context depending on the frame's position in a chunk. 

However, the right-context information is very relevant to learn the acoustic-linguistic attributes of a chunk. Utilizing a modest right and left context may yield improved performance in terms of WER and latency when compared to solely relying on an extensive left-context. Incorporating right-context will thus help to narrow the gap in WER between streaming and non-streaming models. Furthermore, due to the varying temporal resolutions of each layer within the zipformer encoder block, the utilization of right-context frames becomes more efficient. In this work, we deploy chunked masking with right-context as shown in Figure~\ref{fig:mask}(b), where the extent of right-context and left-context can be varied based on requirements. We note that the right-context frames are the frames beyond the chunk boundaries, not within the chunks.


\section{Experiments}\label{sec:experiments}
Below we discuss the database used and experiments conducted to demonstrate the effectiveness of right-context in unified streaming and non-streaming ASR models. 

\subsection{Dataset}\label{sec:dataset}

We conduct experiments using two data setups, using {\it Librispeech} and {\it large in-house conversational data}. In the Librispeech setup, we use the standard 960 hours of training data, as well as test-clean (5.40 hrs) and test-other (5.10 hr) partitions for testing. Using the Librispeech setup we train a conventional conformer transducer streaming model and a baseline zipformer streaming model without any right-context during training. Using this setup, we also train a zipformer streaming model with proposed right-context strategy and a non-streaming model.

Using the {\it large in-house conversational setup}, we train zipformer models without right-context, with right-context and the non-streaming variant.
The in-house training data is derived by combining different open source databases along with in-house conversational and simulated conversational telephonic datasets as shown in Table~\ref{tab:databse_details}. In total we use 12,468 hours of training data. The training data also includes a synthesized corpus generated using a text-to-speech model. We employ diverse in-house test datasets listed in Table~\ref{tab:databse_details} that comprise different domains and accents. The DefinedAI en-in, en-ph, en-au and en-gb subsets correspond to Indian, Filipino, Australian and UK-accented English, respectively. To evaluate the latency and inference time in the server-client setup, we use long conversations as test data to mimic the production use-cases.



\begin{table}[ht]
\caption{Duration and domain information for different training and test sets used in the experiments.}
\label{tab:databse_details} 
\renewcommand{\arraystretch}{1.3}
\resizebox{0.45\textwidth}{!}{
\begin{tabular}{|c c c |}
\hline
\multicolumn{1}{|c|}{Dataset} & \multicolumn{1}{|c|}{Duration (hours)} & \multicolumn{1}{|c|}{Domains}\\
\hline
\hline
\multicolumn{3}{|c|}{Train data}\\
\hline
\hline
\multicolumn{1}{|c|}{Defined AI} & \multicolumn{1}{|c|}{2876.99} & \multicolumn{1}{|c|}{Banking, Insurance, Retail, Telecom}\\
\hline
\multicolumn{1}{|c|}{WoW AI} & \multicolumn{1}{|c|}{5316.76} & \multicolumn{1}{|c|}{Airlines, Auto-insurance, Automotive, Medicare,}\\
\multicolumn{1}{|c|}{} & \multicolumn{1}{|c|}{} & \multicolumn{1}{|c|}{ Customer Service, Home Service, Generic}\\
\hline
\multicolumn{1}{|c|}{Client-1-3} & \multicolumn{1}{|c|}{1457.34} & \multicolumn{1}{|c|}{Telecom}\\
\hline
\multicolumn{1}{|c|}{Client-4} & \multicolumn{1}{|c|}{52.55} & \multicolumn{1}{|c|}{Healthcare}\\
\hline
\multicolumn{1}{|c|}{Client-5} & \multicolumn{1}{|c|}{75.00} & \multicolumn{1}{|c|}{Airlines}\\
\hline
\multicolumn{1}{|c|}{Client-6} & \multicolumn{1}{|c|}{45.42} & \multicolumn{1}{|c|}{Banking}\\
\hline
\multicolumn{1}{|c|}{Client-7} & \multicolumn{1}{|c|}{13.75} & \multicolumn{1}{|c|}{Medicare}\\
\hline
\multicolumn{1}{|c|}{Client-8-16} & \multicolumn{1}{|c|}{956.95} & \multicolumn{1}{|c|}{Generic}\\
\hline
\multicolumn{1}{|c|}{Spgispeech} & \multicolumn{1}{|c|}{866.45} & \multicolumn{1}{|c|}{Generic}\\
\hline
\multicolumn{1}{|c|}{Switchboard} & \multicolumn{1}{|c|}{309.99} & \multicolumn{1}{|c|}{Generic}\\
\hline
\multicolumn{1}{|c|}{CommonVoice} & \multicolumn{1}{|c|}{179.15} & \multicolumn{1}{|c|}{Generic}\\
\hline
\multicolumn{1}{|c|}{GigaSpeech} & \multicolumn{1}{|c|}{124.14} & \multicolumn{1}{|c|}{Generic}\\
\hline
\multicolumn{1}{|c|}{Alphadigits} & \multicolumn{1}{|c|}{30.83} & \multicolumn{1}{|c|}{Alphadigits}\\
\hline
\multicolumn{1}{|c|}{Synthesised data} & \multicolumn{1}{|c|}{162.72} & \multicolumn{1}{|c|}{Generic, Banking}\\
\hline
\hline
\multicolumn{3}{|c|}{Test data}\\
\hline
\hline
\multicolumn{1}{|c|}{Defined AI en-in } & \multicolumn{1}{|c|}{85.34} & \multicolumn{1}{|c|}{Banking, Insurance, Retail, Telecom}\\
\hline
\multicolumn{1}{|c|}{Defined AI en-gb} & \multicolumn{1}{|c|}{52.08} & \multicolumn{1}{|c|}{Banking, Insurance, Retail, Telecom}\\
\hline
\multicolumn{1}{|c|}{Defined AI en-ph} & \multicolumn{1}{|c|}{31.90} & \multicolumn{1}{|c|}{Banking, Insurance, Retail, Telecom}\\
\hline
\multicolumn{1}{|c|}{Defined AI en-au} & \multicolumn{1}{|c|}{51.28} & \multicolumn{1}{|c|}{Banking, Insurance, Retail, Telecom}\\
\hline
\multicolumn{1}{|c|}{Client-1} & \multicolumn{1}{|c|}{12.36} & \multicolumn{1}{|c|}{Telecom}\\
\hline
\multicolumn{1}{|c|}{Client-2} & \multicolumn{1}{|c|}{3.60} & \multicolumn{1}{|c|}{Telecom}\\
\hline
\multicolumn{1}{|c|}{Client-3} & \multicolumn{1}{|c|}{7.64} & \multicolumn{1}{|c|}{Telecom}\\
\hline
\multicolumn{1}{|c|}{Client-17} & \multicolumn{1}{|c|}{35.96} & \multicolumn{1}{|c|}{Generic}\\
\hline
\hline
\multicolumn{3}{|c|}{Latency test data}\\
\hline
\multicolumn{1}{|c|}{Long calls testset} & \multicolumn{1}{|c|}{310.09} & \multicolumn{1}{|c|}{Generic}\\
\hline
\end{tabular}
}
\end{table}

\subsection{Experimental setup}\label{sec:experimental_setup} 
To assess the effectiveness of the proposed approach to unify streaming and non-streaming  ASR models, we setup our experiments using Librispeech and large in-house conversational dataset. For both the setups, we evaluate different baseline and right-context models using Icefall's simulated streaming decoding approach. We further evaluate the large in-house ASR models in server-client production setup.



\subsubsection{Librispeech models}
Using the Librispeech setup, we initially train a baseline conformer transducer streaming model~\cite{kuang2022pruned} (${\text{Conformer}}_{\text{Baseline}}$) without any right-context. Further, we train two zipformer streaming models: the baseline model ($\text{Libri}_{\text{Baseline}}$), the right-context model ($\text{Libri}_{\text{RC-0-64-128-256}}$). Additionally, a non-streaming model (${\text {Libri}}_{\text{NS}}$) is trained using this setup.

\subsubsection{Large-data conversational models}
Utilizing the large in-house conversational English data, we showcase the efficacy of the proposed approach in a more challenging conversational environment with different test cases comprising different domains and accents. Using this data, We train two streaming zipformer models: $\text{Large}_{\text{Baseline}}$, 
 and $\text{Large}_{\text{RC-0-64-128-256}}$, and a non-streaming model ${\text {Large}}_{\text{NS}}$. 

\subsubsection{Training setup}
All experiments described above (except ${\text{Conformer}}_{\text{Baseline}}$ model) adhere to the standard {\it zipformer} recipe\footnote{https://tinyurl.com/2whxxub2} within the Icefall toolkit. The conformer model (${\text{Conformer}}_{\text{Baseline}}$) is trained using the  ${\text{pruned\_transducer\_stateless4}}$ recipe in Icefall toolkit. We use the zipformer-medium setup for Librispeech model and zipformer-large for the large in-house models~\cite{yao2023zipformer}.
The base learning rate is 0.045 for the Librispeech setup, and 0.05 for the large in-house model training. Additionally, the chunk-size varies among the values [16, 32, 64] frames during training, where, each frame corresponds to 10 ms in both training and decoding. Based on our experiments on a a small-data setup, we use varying numbers of right-context frames by randomly choosing from the set  $ \{ 0, 64, 128, 256\}$  for each batch during training. All models undergo training for up to 30 epochs, using eight NVIDIA V100 GPUs. 

Evaluation is conducted using 128 left-context frames, a chunk size of 32 frames, 30 epochs with an averaging over 6. We evaluate different baseline and right-context models using Icefall's simulated streaming decoding approach for both Librispeech and Large in-house setups. We also demonstrate performance in server-client setup for the in-house models.

\subsubsection{Server-client-based evaluation}
To demonstrate the performance of the proposed unified ASR training approach, we evaluate the in-house models ($\text{Large}_{\text{Baseline}}$, $\text{Large}_{\text{RC-0-64-128-256}}$, ${\text {Large}}_{\text{NS}}$) in server-client setup. We use Sherpa websocket server for real-time streaming\footnote{https://github.com/k2-fsa/sherpa}. The ASR model is loaded on a cpp-based websocket server, which listens to a specific port on a server machine. A Python client is used to create multiple and simultaneous websocket connections to the server to support concurrent processing. The client streams audio chunks of 500\,ms in real time. When an endpoint is reached in the audio, the transcripts are sent back to the client. ``Final-chunk latency'' is the metric used to measure the latency of the ASR output: latency is measured in the client as the time from when the last chunk is streamed to the server to the time when the final transcript is received back in the client. The server used in this experiment is a g5.2xlarge AWS instance, which has 1 Nvidia A10G GPU, 8 vPUs and 32GB RAM.

\subsection{Evaluation metrics}
We use word error rate ({\it WER}) as the performance metric for recognition accuracy.
Final-chunk latency as described above is evaluated in the client-server setting and simply referred to as {\it {latency}} here.
Another measure to analyze the inference time is inverse real time factor (RTFX).  {\it {RTFX}} is calculated as, $\text{RTFX}=\frac{\text{duration of testset}}{\text{inference time}}$. Higher RTFX corresponds to less inference time. As in production environment, we process multiple calls at the same time, we analyse the latency and RTFX over different concurrency values. Concurrency can be defined as the number of concurrent calls being sent from the client to the server at a given point in time.

We measure latency only for streaming ASR and RTFX for both streaming and non-streaming ASR models. Non-streaming models do not support concurrency in our setup, as they process a conversation by splitting it into smaller segments.

\section{Results}\label{sec:results}
\subsection{Librispeech setup}

\begin{figure}[t]
 \centering
\centerline{\epsfig{figure=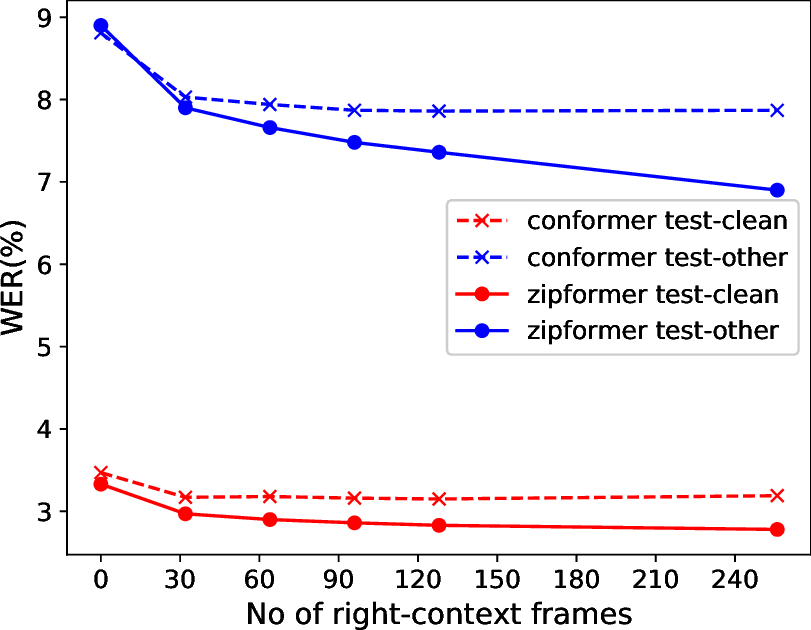,scale=0.4}}
\caption{Comparison of conventional conformer (${\text{Conformer}}_{\text{Baseline}}$) and zipformer ($\text{Medium}_{\text{Baseline}}$) models in terms of WER(\%) with different number of right-context frames during inference.}
\label{fig:conformer_vs_zipformer}
\vspace{-0.3cm}
\end{figure}

In Figure~\ref{fig:conformer_vs_zipformer}, we compare the WER (\%) of the ${\text{Conformer}}_{\text{Baseline}}$ model with that of the zipformer-based $\text{Medium}_{\text{Baseline}}$ model for Librispeech test-clean and test-other testsets. We note that these two models are not trained with right-context. Figure~\ref{fig:conformer_vs_zipformer}, illustrates that during inference, increasing the number of right-context frames leads to WER improvement for both models. However, the zipformer-based model shows more pronounced improvement in WER compared to the conformer model. The enhanced performance is due to the varying frame rates across different encoder blocks in the zipformer architecture, making it a superior choice for a unified ASR model.

In Table~\ref{tab:medium-setup-results}, we show the WER(\%) for $\text{Libri}_{\text{Baseline}}$ and $\text{Libri}_{\text{RC-0-64-128-256}}$  models for different numbers of right-context frames during decoding. A noteworthy observation is the improvement in WER of the $\text{Libri}_{\text{Baseline}}$ model, which decreases from 3.33\% to 2.83\% as the number of decoding right-context frames increases from 0 to 256, despite this model not being trained with right-context. In the baseline model, although we do not explicitly impose right-context frames, the initial frames of a chunk see the entire chunk length as right-context, whereas the later frames do not have access to any right-context.
 The $\text{Libri}_{\text{RC-0-64-128-256}}$ model achieves WERs of 2.43\% in test-clean and 6.55\% in test-other, compared to the baseline model's respective WERs of 3.33\% and 8.90\%, bringing it closer to the non-streaming model's ($\text{Libri}_{\text{NS}}$) performance, as shown in Table~\ref{tab:medium-setup-results}. Across all models, increasing the number of decoding right-context frames consistently contributes to obtaining a viable unified model for both streaming and non-streaming applications.

\begin{table}[ht]
\caption{WER(\%) of the models trained on 960 hours of Librispeech data, including $\text{Libri}_{\text{Baseline}}$, $\text{Libri}_{\text{RC-0-64-128-256}}$ and non-streaming model.}
\label{tab:medium-setup-results} 
\renewcommand{\arraystretch}{1}
\resizebox{0.47\textwidth}{!}{
\begin{tabular}{|c c c c c c|}
\hline
\multicolumn{1}{|c|}{models$\rightarrow$} & \multicolumn{2}{|c|}{$\text{Libri}_{\text{Baseline}}$} & \multicolumn{2}{|c|}{$\text{Libri}_{\text{RC-0-64-128-256}}$}  & \multicolumn{1}{|c|}{\multirow{2}{*}{$\text{Libri}_{\text{NS}}$}}\\
\cline{1-5}
\multicolumn{1}{|c|}{\#Decoding RC frames$\downarrow$} & \multicolumn{1}{|c|}{test-clean} & \multicolumn{1}{|c|}{test-other} & \multicolumn{1}{|c|}{test-clean} & \multicolumn{1}{|c|}{test-other} & \multicolumn{1}{|c|}{}\\
\hline
\hline
\multicolumn{1}{|c|}{0} & \multicolumn{1}{|c|}{3.33} & \multicolumn{1}{|c|}{8.90} & \multicolumn{1}{|c|}{4.43} & \multicolumn{1}{|c|}{9.50} & \multicolumn{1}{|c|}{}\\
\cline{1-5}
\multicolumn{1}{|c|}{32} & \multicolumn{1}{|c|}{2.97} & \multicolumn{1}{|c|}{7.90} & \multicolumn{1}{|c|}{2.80} & \multicolumn{1}{|c|}{7.10} & \multicolumn{1}{|c|}{test-clean: 2.38}\\
\cline{1-5}
\multicolumn{1}{|c|}{64} & \multicolumn{1}{|c|}{2.90} & \multicolumn{1}{|c|}{7.66} & \multicolumn{1}{|c|}{2.74} & \multicolumn{1}{|c|}{6.89} & \multicolumn{1}{|c|}{test-other: 5.72}\\
\cline{1-5}
\multicolumn{1}{|c|}{96} & \multicolumn{1}{|c|}{2.86} & \multicolumn{1}{|c|}{7.48} & \multicolumn{1}{|c|}{2.58} & \multicolumn{1}{|c|}{6.85} & \multicolumn{1}{|c|}{}\\
\cline{1-5}
\multicolumn{1}{|c|}{128} & \multicolumn{1}{|c|}{2.83} & \multicolumn{1}{|c|}{7.36} & \multicolumn{1}{|c|}{2.46} & \multicolumn{1}{|c|}{6.70} & \multicolumn{1}{|c|}{}\\
\cline{1-5}
\multicolumn{1}{|c|}{256} & \multicolumn{1}{|c|}{2.81} & \multicolumn{1}{|c|}{7.36} & \multicolumn{1}{|c|}{2.43} & \multicolumn{1}{|c|}{6.55} & \multicolumn{1}{|c|}{}\\
\hline
\end{tabular}
}
\end{table}

\subsection{Large in-house conversational setup}



In Table~\ref{tab:large-setup-results}, we depict the WER values of the  $\text{Large}_{\text{Baseline}}$ and $\text{Large}_{\text{RC-0-64-128-256}}$ models with the number of right-context frames in decoding varying from 0 to 256. We can observe that the WER of $\text{Large}_{\text{Baseline}}$ improves as we increase the number of right-context frames in decoding, although the model is not trained with right-context. However, the right-context training strategy presented in this paper helps to further improve the performance of the $\text{Large}_{\text{RC-0-64-128-256}}$ model across all testsets. Notably, with 64 right-context frames during decoding, the average WER improves to 8.31\% compared to 10.34\% in the baseline without right context during training and decoding. Moreover, the results in  Table~\ref{tab:large-setup-results} exhibit the convergence of the streaming model's performance towards the non-streaming model with the proposed right-context attention mask. This convergence signifies the potential for deploying a streaming ASR model in place of its corresponding non-streaming counterpart, facilitated by increasing the decoding right context frames. Ultimately, these results affirm that a unified zipformer-based model can effectively serve both streaming and non-streaming applications through the proposed right-context chunked and hybrid attention masking training methods. Apart from unifying streaming and non-streaming models, the proposed approach adds flexibility to choose a balance between accuracy and latency by selecting an suitable number of right-context frames in decoding according to requirement. 

\begin{table}[ht]
\caption{WER(\%) of the models trained on 12,460 hours of in-house conversational data, including $\text{Large}_{\text{Baseline}}$, $\text{Large}_{\text{RC-0-64-128-256}}$, and non-streaming model with in-house testsets.}
\label{tab:large-setup-results} 
\renewcommand{\arraystretch}{1.2}
\resizebox{0.47\textwidth}{!}{
\begin{tabular}{|c c c c c c c c c|}
\hline
\multicolumn{1}{|c|}{Model$\rightarrow$} & \multicolumn{5}{|c|}{$\text{Large}_{\text{Baseline}}$} & \multicolumn{2}{|c|}{$\text{Large}_{\text{RC-0-64-128-256}}$} & \multicolumn{1}{|c|}{\multirow{2}{*}{$\text{Large}_{\text{NS}}$}}\\
\cline{1-8}
\multicolumn{1}{|c|}{\#Decoding RC frames$\rightarrow$} & \multicolumn{1}{|c|}{0} & \multicolumn{1}{|c|}{32} & \multicolumn{1}{|c|}{64}   & \multicolumn{1}{|c|}{128}  & \multicolumn{1}{|c|}{256} & \multicolumn{1}{|c|}{32} & \multicolumn{1}{|c|}{64} & \multicolumn{1}{|c|}{}\\
\hline
\hline
\multicolumn{1}{|c|}{Defined AI en-au} & \multicolumn{1}{|c|}{6.95} & \multicolumn{1}{|c|}{6.75} & \multicolumn{1}{|c|}{6.72}   & \multicolumn{1}{|c|}{6.72} & \multicolumn{1}{|c|}{6.72} & \multicolumn{1}{|c|}{6.41}  & \multicolumn{1}{|c|}{6.39} & \multicolumn{1}{|c|}{6.2}\\
\hline
\multicolumn{1}{|c|}{Defined AI en-in} & \multicolumn{1}{|c|}{6.28}  & \multicolumn{1}{|c|}{6.01}  & \multicolumn{1}{|c|}{5.96} & \multicolumn{1}{|c|}{5.92}  & \multicolumn{1}{|c|}{5.90} & \multicolumn{1}{|c|}{5.80} & \multicolumn{1}{|c|}{5.76} & \multicolumn{1}{|c|}{5.7}\\
\hline
\multicolumn{1}{|c|}{Defined AI en-ph} & \multicolumn{1}{|c|}{7.21} & \multicolumn{1}{|c|}{6.82} & \multicolumn{1}{|c|}{6.75}  & \multicolumn{1}{|c|}{6.69} & \multicolumn{1}{|c|}{6.68} & \multicolumn{1}{|c|}{6.29}  & \multicolumn{1}{|c|}{6.29}  & \multicolumn{1}{|c|}{7.9}\\
\hline
\multicolumn{1}{|c|}{Defined AI en-gb} & \multicolumn{1}{|c|}{5.80} & \multicolumn{1}{|c|}{5.42} & \multicolumn{1}{|c|}{5.40}  & \multicolumn{1}{|c|}{5.38} & \multicolumn{1}{|c|}{5.33} & \multicolumn{1}{|c|}{4.72}  & \multicolumn{1}{|c|}{4.73}  & \multicolumn{1}{|c|}{4.5}\\
\hline
\multicolumn{1}{|c|}{Client-1} & \multicolumn{1}{|c|}{13.81} & \multicolumn{1}{|c|}{12.85} & \multicolumn{1}{|c|}{12.69} & \multicolumn{1}{|c|}{12.74}  & \multicolumn{1}{|c|}{12.8}  & \multicolumn{1}{|c|}{10.9} & \multicolumn{1}{|c|}{10.74}  & \multicolumn{1}{|c|}{10.5}\\
\hline
\multicolumn{1}{|c|}{Client-2} & \multicolumn{1}{|c|}{15.66} & \multicolumn{1}{|c|}{14.02} & \multicolumn{1}{|c|}{13.88} & \multicolumn{1}{|c|}{13.91}  & \multicolumn{1}{|c|}{14.00} & \multicolumn{1}{|c|}{11.88} & \multicolumn{1}{|c|}{11.60} & \multicolumn{1}{|c|}{11.1}\\
\hline
\multicolumn{1}{|c|}{Client-3} & \multicolumn{1}{|c|}{13.64} & \multicolumn{1}{|c|}{12.83} & \multicolumn{1}{|c|}{12.63} & \multicolumn{1}{|c|}{12.53}  & \multicolumn{1}{|c|}{12.48} & \multicolumn{1}{|c|}{11.05}  & \multicolumn{1}{|c|}{10.91}  & \multicolumn{1}{|c|}{10.4}\\
\hline
\multicolumn{1}{|c|}{Client-17} & \multicolumn{1}{|c|}{13.38} & \multicolumn{1}{|c|}{12.08} & \multicolumn{1}{|c|}{11.62} & \multicolumn{1}{|c|}{11.42}  & \multicolumn{1}{|c|}{11.28} & \multicolumn{1}{|c|}{10.60}  & \multicolumn{1}{|c|}{10.08} & \multicolumn{1}{|c|}{9.8}\\
\hline
\multicolumn{1}{|c|}{\bf{Average}} & \multicolumn{1}{|c|}{\bf 10.34} & \multicolumn{1}{|c|}{\bf 9.50} & \multicolumn{1}{|c|}{\bf 9.45} & \multicolumn{1}{|c|}{\bf 9.41}  & \multicolumn{1}{|c|}{\bf 9.30} & \multicolumn{1}{|c|}{\bf 8.45}  & \multicolumn{1}{|c|}{\bf 8.31} & \multicolumn{1}{|c|}{\bf 8.26}\\
\hline
\end{tabular}
}
\end{table}


\subsubsection{Server-client setup}
As discussed in Section~\ref{sec:experimental_setup}, we deploy the large in-house conversation model in server-client environment. In Table~\ref{tab:onnx-wer-large-setup-results}, we show the WERs for the $\text{Large}_{\text{Baseline}}$ model with no right-context in decoding and the $\text{Large}_{\text{RC-0-64-128-256}}$ model with 0, 32, and 64 right-context frames in decoding along with the non-streaming model ($\text{Large}_{\text{NS}}$). We note that for the same model there is a difference in performance between the simulated streaming and real streaming (server-client) environments, because of the padding involved in the real streaming case. However, from Table~\ref{tab:onnx-wer-large-setup-results} we can observe that the average WER of the in-house model improves from  9.0\% to 8.2\% with the streaming model, approaching the non-streaming model.

\begin{table}[h]
\caption{WER(\%) of the $\text{Large}_{\text{RC-0-64-128-256}}$ and non-streaming models trained on 12,460 hours of in-house conversational data for different in-house testsets.}
\label{tab:onnx-wer-large-setup-results} 
\renewcommand{\arraystretch}{1}
\resizebox{0.45\textwidth}{!}{
\begin{tabular}{|c c c c c|}
\hline
\multicolumn{1}{|c|}{Model$\rightarrow$} & \multicolumn{3}{|c|}{$\text{Large}_{\text{RC-0-64-128-256}}$} & \multicolumn{1}{|c|}{\multirow{2}{*}{$\text{Large}_{\text{NS}}$}}\\
\cline{1-4}
\multicolumn{1}{|c|}{\#Decoding RC frames$\rightarrow$}  & \multicolumn{1}{|c|}{0} & \multicolumn{1}{|c|}{32} & \multicolumn{1}{|c|}{64}  & \multicolumn{1}{|c|}{}\\
\hline
\hline
\multicolumn{1}{|c|}{Defined AI en-au} & \multicolumn{1}{|c|}{6.5} & \multicolumn{1}{|c|}{6.3} & \multicolumn{1}{|c|}{6.2} & \multicolumn{1}{|c|}{6.2}\\
\hline
\multicolumn{1}{|c|}{Defined AI en-in} & \multicolumn{1}{|c|}{6.0} & \multicolumn{1}{|c|}{5.7} & \multicolumn{1}{|c|}{5.3}  &  \multicolumn{1}{|c|}{5.7}\\
\hline
\multicolumn{1}{|c|}{Defined AI en-ph} & \multicolumn{1}{|c|}{9.9} & \multicolumn{1}{|c|}{9.5} & \multicolumn{1}{|c|}{8.5}   & \multicolumn{1}{|c|}{7.9}\\
\hline
\multicolumn{1}{|c|}{Defined AI en-gb} & \multicolumn{1}{|c|}{5.0} & \multicolumn{1}{|c|}{4.7}  & \multicolumn{1}{|c|}{4.2} & \multicolumn{1}{|c|}{4.5}\\
\hline
\multicolumn{1}{|c|}{Client-1} & \multicolumn{1}{|c|}{11.3} & \multicolumn{1}{|c|}{10.4} & \multicolumn{1}{|c|}{10.4}  & \multicolumn{1}{|c|}{10.5}\\
\hline
\multicolumn{1}{|c|}{Client-2} & \multicolumn{1}{|c|}{12.1} & \multicolumn{1}{|c|}{11.2} & \multicolumn{1}{|c|}{11.0}  & \multicolumn{1}{|c|}{11.1}\\
\hline
\multicolumn{1}{|c|}{Client-3}  & \multicolumn{1}{|c|}{10.9} & \multicolumn{1}{|c|}{10.4} & \multicolumn{1}{|c|}{10.4} & \multicolumn{1}{|c|}{10.4}\\
\hline
\multicolumn{1}{|c|}{Client-17} & \multicolumn{1}{|c|}{10.7} & \multicolumn{1}{|c|}{10.9} & \multicolumn{1}{|c|}{9.8} & \multicolumn{1}{|c|}{9.8}\\
\hline
\multicolumn{1}{|c|}{\bf Average} & \multicolumn{1}{|c|}{\bf 9.0} & \multicolumn{1}{|c|}{\bf 8.5} & \multicolumn{1}{|c|}{\bf 8.2} & \multicolumn{1}{|c|}{\bf 8.2}\\
\hline
\end{tabular}
}
\end{table}

\begin{table}[h]
\caption{Latency (sec) and RTFX values of the $\text{Large}_{\text{RC-0-64-128-256}}$ and  $\text{Large}_{\text{NS}}$  models trained on 12,460 hours of in-house conversational data in server-client setup for the long calls testset.}
\label{tab:onnx-latency-large-setup-results} 
\renewcommand{\arraystretch}{1.2}
\resizebox{0.47\textwidth}{!}{
\begin{tabular}{|c c c c c c c c c|}
\hline
\multicolumn{1}{|c|}{Model$\rightarrow$}  & \multicolumn{6}{|c|}{$\text{Large}_{\text{RC-0-64-128-256}}$} & \multicolumn{1}{|c|}{\multirow{2}{*}{$\text{Large}_{\text{NS}}$}}\\
\cline{1-7}
\multicolumn{1}{|c|}{\#Decoding RC frames$\rightarrow$}  & \multicolumn{2}{|c|}{0} & \multicolumn{2}{|c|}{32} & \multicolumn{2}{|c|}{64}  & \multicolumn{1}{|c|}{}\\
\cline{1-7}
\multicolumn{1}{|c|}{Concurrency$\downarrow$}  & \multicolumn{1}{|c|}{Latency} & \multicolumn{1}{|c|}{RTFX} & \multicolumn{1}{|c|}{Latency}  & \multicolumn{1}{|c|}{RTFX}  & \multicolumn{1}{|c|}{Latency}  & \multicolumn{1}{|c|}{RTFX} & \multicolumn{1}{|c|}{}\\
\hline
\hline
\multicolumn{1}{|c|}{100} & \multicolumn{1}{|c|}{1.41} & \multicolumn{1}{|c|}{82.65} & \multicolumn{1}{|c|}{1.44}  & \multicolumn{1}{|c|}{82.66} & \multicolumn{1}{|c|}{1.47}  & \multicolumn{1}{|c|}{82.66} & \multicolumn{1}{|c|}{}\\
\cline{1-7}
\multicolumn{1}{|c|}{200} & \multicolumn{1}{|c|}{2.17} & \multicolumn{1}{|c|}{163.76} & \multicolumn{1}{|c|}{2.17} & \multicolumn{1}{|c|}{163.77}  & \multicolumn{1}{|c|}{2.35}  & \multicolumn{1}{|c|}{163.73} & \multicolumn{1}{|c|}{143.89}\\
\cline{1-7}
\multicolumn{1}{|c|}{300} & \multicolumn{1}{|c|}{2.45} & \multicolumn{1}{|c|}{242.27} & \multicolumn{1}{|c|}{2.83}  & \multicolumn{1}{|c|}{242.21} & \multicolumn{1}{|c|}{3.24} & \multicolumn{1}{|c|}{242.23} & \multicolumn{1}{|c|}{}\\
\hline
\end{tabular}
}
\end{table}
Apart from WER, latency or inference time plays a crucial role in industrial streaming ASR models. In Table~\ref{tab:onnx-latency-large-setup-results}, we show the latency and RTFX values of the $\text{Large}_{\text{RC-0-64-128-256}}$ model for different numbers of decoding right-context frames for concurrency of 100, 200 and 300. In this evaluation we use the long conversations testset in Table~\ref{tab:databse_details}. From Table~\ref{tab:onnx-latency-large-setup-results}, we can observe that there is no significant degradation of user-perceived latency as right-context increases. The RTFX values of the streaming $\text{Large}_{\text{RC-0-64-128-256}}$ model are higher than that of the non-streaming model in all the cases. The greater RTFX demonstrates less inference time for the streaming model with right-context compared to the non-streaming model. For the streaming application, with the introduction of right-context we observe increase in accuracy  and a small degradation in latency; for the non-streaming use-case the accuracy drops with the reduction in inference time or latency. As we further increase the number of decoding right-context frames, the accuracy of streaming model eventually comes close to that of the non-streaming model.

\section{Conclusions}\label{sec:summary}
We propose to unify streaming and non-streaming zipformer ASR models by incorporating right-context frames. We employ a chunked attention masking strategy with dynamic right-context to improve the WER of a zipformer-based streaming ASR model. We observe that baseline streaming models trained without right-context eventually shows improved performance with right-context during inference. With the increase in decoding right-context frames, the gap in WER\% between the streaming and non-streaming model decreases, thereby validating the proposed unified training of streaming and non-streaming zipformer models. Our approach yields a flexible ASR model that can achieve the desired accuracy-latency tradeoff during inference, based on application requirements. 

\bibliography{mybib}

\newpage
\appendix
\section{Experiments to find optimal right-context training setup}

\begin{figure}[h]
 \centering
\centerline{\epsfig{figure=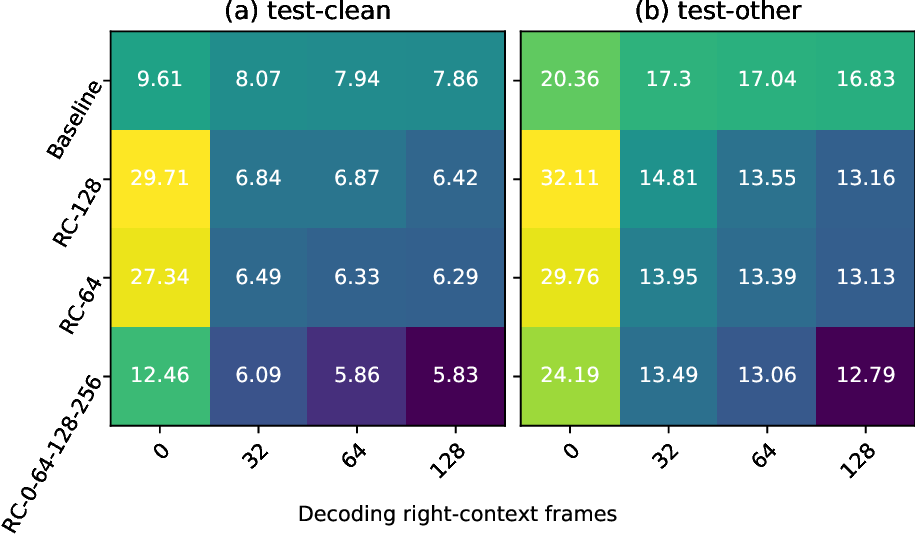,scale=0.5}}
{\caption{\footnotesize{WER(\%) of the models trained on 100 hours of clean Librispeech training data, varying the number of right-context frames, evaluated on (a) test-clean and (b) test-other datasets.}}
\label{fig:smallsetup-test-clean}}
\end{figure}

To refine the number of right-context frames that the model acquires during the training process, we first train various zipformer models using the small-scale Librispeech 100 hours training dataset.

 First, we develop a baseline streaming model without right-context.  Subsequently, we train different models: with constant 128 frames of right-context in training ({\it {RC-128}}), and another incorporating 64 frames of right-context, termed as the {\it{RC-64}} model. In successive models, we introduce variability in the number of right-context frames utilized during training. Specifically, within each batch, the number of right-context frames is randomly selected from the set $\{ 0, 64, 128, 256 \}$  for the {\it{RC-0-64-128-256}} model. 
In these models the number of right-context frames is constant over the training. We note that the duration of contexts of {\it{RC-64}} and {\it{RC-128}} are $1.28sec$ and $2.56sec$, respectively.

To assess the impact of the number of right-context frames used in decoding, we evaluate each model for 0, 32, 64, and 128 right-context frames. 

We found that all models trained with right-context outperform the baseline model without right-context. Notably, models trained with varying right-context frames during training demonstrate superior performance compared to those trained with fixed right-context frames.  Among these, the RC-0-64-128-256 model achieves the lowest WER. In all cases, increasing the number of right-context frames in decoding leads to improved performance. Additionally, we note that models trained with right-context experience degraded performance when decoded without right-context frames. Figure~\ref{fig:smallsetup-test-clean} (a) and Figure~\ref{fig:smallsetup-test-clean} (b) show the WER values corresponding to the test-clean and test-other testsets, respectively. In Figure~\ref{fig:smallsetup-test-clean}(a), we observe a diagonal improvement in WER from 9.61\% to 5.83\% with the introduction of right context. A similar trend is evident in Figure~\ref{fig:smallsetup-test-clean}(b) for the test-other testset.

\end{document}